\documentstyle[tighten,preprint,eqsecnum,aps,prd]{revtex}

\def\a{\alpha}
\def\b{\beta}
\def\e{\epsilon}
\def\g{\gamma}
\def\l{\lambda}
\def\p{\phi}
\def\t{\theta}

\def\be{\begin{equation}}
\def\bea{\begin{eqnarray}}
\def\eea{\end{eqnarray}}
\def\ee{\end{equation}}
\def\bi{\begin{itemize}}
\def\ei{\end{itemize}}

\def\d{\delta}

\begin{document}
\draft

\preprint{\vbox{\baselineskip=12pt
\rightline{IUCAA-2/2000}
\rightline{}
\rightline{gr-qc/0002010}
}}
\title{Detection of gravitational waves from inspiraling compact binaries using 
a network of interferometric detectors\footnote{Based on talk given at GWDAW-99,
Rome, in December 1999.}}
\author{
Sukanta Bose\footnote{Electronic address:
{\em boses@cf.ac.uk}}(1), Archana Pai\footnote{Electronic
address: {\em apai@iucaa.ernet.in}}(2), and Sanjeev V. Dhurandhar\footnote{
Electronic address: {\em sanjeev@iucaa.ernet.in}}(2,3)}
\address{(1) Department of Physics and Astronomy, P. O. 
Box 913, Cardiff University, CF24 3YB, United Kingdom}
\address{(2) Inter-University Centre for Astronomy and Astrophysics, Post Bag 4,
Ganeshkhind,\\ Pune 411007, India}
\address{(3) Max Planck Institut fur Gravitationphysik,
Albert Einstein Institut, Am Muhlenberg 1, Golm, D-14476, Germany}

\maketitle
\begin{abstract}

We formulate the data analysis problem for the detection of the 
Newtonian waveform from an inspiraling compact-binary by a network of 
arbitrarily oriented and arbitrarily distributed laser interferometric 
gravitational wave detectors. We obtain for the first time the relation 
between the optimal statistic and the magnitude of the network correlation
vector, which is constructed from the matched network-filter. This generalizes
the calculation reported in an earlier work (gr-qc/9906064), where the 
detectors are taken to be coincident.

\end{abstract}

\narrowtext

\vfil
\pagebreak

\section{Introduction}	
\label{sec:intro}

Inspiraling compact binaries form prime candidates for detection by 
earth-based interferometric gravitational-wave (GW) detectors owing to the 
well understood waveform (chirp) emitted by them. Searching for chirps using 
a network of such detectors is gaining importance due to (a) its superior 
sensitivity {\it vis a vis} that of a constituent detector [1] and (b) 
improving feasibility for a real-time computational search. Here, we formulate
the problem of how to optimally detect the Newtonian chirp using a network of 
arbitrarily orientated and arbitrarily located detectors. This extends a 
similar study in Ref. [2] of a network with coincident detectors.

We use the maximum likelihood method for optimizing the detection problem.[3]
A single likelihood ratio (LR) is deduced for the entire network. A 
super-threshold value for the maximized likelihood ratio (MLR) implies a
detection. The MLR is obtained by maximizing the LR over the eight parameters
that determine the Newtonian chirp: the distance $r$ to the binary, 
the inital phase  $\delta$ of the waveform, the polarization angle 
$\psi$, the inclination angle $\epsilon$ of the binary orbit, the time of 
arrival $t_a$ at a fiducial detector (fide),
the source-direction angles $\{\phi, \theta\}$, and the chirp time $\xi$.
In principle, this can always be done numerically using 
a grid in the eight dimensional parameter space. In practice, however, such a 
strategy is computationally unfeasible and wasteful. We show that maximization
of the LR over four parameters, $\{r,\delta,\psi,\epsilon\}$,
can be performed analytically using the symmetries in detector responses.
This allows us to scan this parameter subspace {\em continuously}. 
Further, the Fast Fourier 
Transform (FFT) can be used to maximize LR over $t_a$, as in the case of a single 
detector. Such an analytic maximization and the FFT allow us to save 
substantially on computational costs.
 Numerical maximization is 
required over the remaining parameters, $\{\phi, \theta, \xi\}$, which we 
discuss in a future work. Here, we follow the convention laid out in 
Ref. [2].

\section{The signal}

There are four distinctly different reference frames of interest, associated 
with the source, wave, fide, and a representative detector in the network. 
Physical quantities in these frames are related by orthogonal transformations,
${\cal O}_{\rm k}$, which are defined in
terms of three sets of Euler angles that specify the orientation of one frame
with respect to another.[4] Let ${\sf x}$ be an arbitrary 
three-dimensional real vector. Then,
${\sf x}_{\rm wave} = {\cal O}_1(\psi, \epsilon, 0) {\sf x}_{\rm source}$,
${\sf x}_{\rm fide} = {\cal O}_2(\p, \t, 0) {\sf x}_{\rm wave}$, and
${\sf x}_{\rm detector} = {\cal O}^{-1}_3(\a, \b, \g) {\sf x}_{\rm fide}$,
Here, the source axes have been chosen in accordance with Ref. [5].

The wave tensor $w_{ij}$ associated with any source
can be expanded in terms of the STF-2 tensors ${\cal Y}^{ij}_{2n}$ in an 
arbitrary frame as [2]:
\be \label{wstf2a}
w^{ij}(t) = {\sqrt{ 2\pi\over 15}} \left[ \left(h_+(t) - ih_\times (t)\right)
T_{2}{}^n {\cal Y}^{ij}_{2n} + \left(h_+(t) + ih_\times (t)\right)
T_{-2}{}^n {\cal Y}^{ij}_{2n} \right] \ \ ,
\ee
where $i$, $j$ denote spatial indices, and $h_+$ and $h_\times$ are the two 
GW polarizations in the transverse-traceless gauge, as measured in some 
given frame. The expansion coefficients $T_{\pm 2}{}^n$ are the Gel'fand 
functions,[2] which depend on the Euler angles through which one must rotate 
that frame into the frame in which $w^{ij}$ is being analyzed.
The above form suggests the definitions,
$e_L^{ij} = \sqrt{8\pi /15}\>T_{2}{}^n  {\cal Y}^{ij}_{2n}$ and
$e_R^{ij} = \sqrt{8\pi /15}\>T_{-2}{}^n  {\cal Y}^{ij}_{2n}$,
for the left- and right-circular polarization tensors, respectively. They 
obey, $e_L^{ij\>*} = e_R^{ij}$, $e_{L,\> R}^{ij} \>e_{L,\> R\> ij}^* =1$, 
and $e_{L,\> R}^{ij} \>e_{R,\>L\> ij}^* = 0$, in any frame.
Thus,
\be\label{wS}
w^{ij} (t)= {\rm Re} \left[ \left(h_+(t) + ih_\times (t)
\right) e_R^{ij}\right] \equiv 2\kappa \> {\rm Re} \left[ 
R(t) e_R^{ij}\right] \ \ ,
\ee
where 
$\kappa ={\sqrt \xi} / r$ (up to a normalization factor) and 
$R\equiv (h_+(t) + ih_\times (t))/(2\kappa)$. For a chirp, we define $R$ in 
the source frame.[5] Then $h_{+,\times}$ are GW amplitudes for a face-on 
binary (i.e., for $\e=0$), and $R$ depends only on $\{\d,t_a,\xi\}$.

The response amplitude (i.e., the signal) in the $I$-th detector is the scalar 
product $s^I = w^{ij} d_{ij}^I$, which depends on projections of $e_{L,\> R}^{
ij}$ onto the $I$-th detector tensor, $d_{ij}^I$. One such projection defines 
the extended beam-pattern function: 
\be \label{Fdef}
F^I = e_L^{ij}d_{ij}^I \equiv  T_{2}{}^p (\psi, 
\epsilon, 0)  {D}_p^I \>,\quad p=\pm2 \ \ ,
\ee
which corresponds to the left-circular polarization. Above,
\be\label{LI}
{D}^{I}_{p}\equiv\sqrt{8\pi\over 15} \> T_{p}{}^n (\p,\t,0)d^{I}_{ij} 
{\cal Y}^{ij}_{2n} =ig^I T_{p}{}^n (\p, \t, 0)
\left(T^{I*}_{2n} - T^{I*}_{-2n} \right) \ \ ,
\ee
where $T^{I}_{\pm 2n} =T_{\pm 2n}(\a_{I},\b_{I},\g_{I})$ and $d_{ij}^I=g^I
(n_{1i}^I n_{1j}^I -n_{2i}^In_{2j}^I)$, with ${\bf n}^I_{1,2}$ being unit 
vectors along the two arms of the $I$-th interferometer, respectively. Also,
$g^I$ is the detector's noise power spectral density.[6]  Then,
\be  \label{sigW}
s^{I}(t)= 2 \kappa \>{\rm Re} \left( F^{I*} 
R^{I} \right) \equiv 2 \kappa \>{\rm Re} \left( F^{I*} 
S^{I} e^{i\delta}\right)
\ee 
where 
$R^{I}$ is defined via Eq. (\ref{wS}) and $S^{I}$ is independent of $\d$. 

\section{The optimal network statistic}
\label{sec:detect}

Under the Neyman-Pearson decision criterion,[3] the optimal network statistic
is the network LR, $\l$. If the noise in each detector is additive and 
independent of the noise in any 
other detector in the network, then $\l$ reduces to a product of the individual
detector LR's.[2] Further, for Gaussian noise,[7] the logarithmic likelihood
ratio (LLR), $\ln \lambda$, simplifies to the following sum of LLR's  of $N$ 
individual detectors [2]:
\be
\ln\lambda = \sum_{I=1}^N \langle s_{I}, x_{I} \rangle_{I} -
{1 \over 2}\sum_{I=1}^N \langle s_{I}, s_{I} \rangle_{I} 
= {\sf b} \sum_{I=1}^N \langle z_{I}, x_{I}\rangle_{I} - 
{1\over 2} {\sf b}^2 \ \ , \label{LRb}
\ee
where 
${\sf b}\equiv 2\kappa (\sum_{I=1}^N \|F_{I}\|^2)^{1/2}$
and $z_{I} = s_{I}/{\sf b}$.  Above, 
$r$ appears only in ${\sf b}$.

Maximizing $\ln\lambda$ with respect to ${\sf b}$ and $\d$ gives,
$\ln \lambda |_{\hat{\sf b},\hat{\d}} =
\left|\sum_{I=1}^N Q_{I} C_{I}^*\right|^2 /2$,
where 
$Q_{I} \equiv 2\kappa F_{I}/{\sf b}$ and $C_{I}^* \equiv \langle S_{I}, x_{I}
\rangle_{I}$.[2] 
This shows that the network vector ${\sf S}$, with $S_{I}$'s as its 
components, is the matched network-filter. Also, $\ln \lambda |_{\hat{\sf b},
\hat{\d}}$ is a function of six parameters, namely,
$\{\psi,\e,t_a,\p,\t,\xi\}$. To extend these results to the case where 
the detectors are arbitrarily located, note that the
dependence of $\ln \lambda |_{\hat{\sf b},\hat{\d}}$ on $\{\psi, \epsilon\}$ 
can be isolated. This is because the network vector ${\sf Q}$, with $Q^I$'s as
its components, is:
\be
{\sf Q} =  {\parallel {\sf F} \parallel}^{-1}
\left( T_{2}^{-2} (\psi,\e, 0) {\sf D}_{-2}
+  T_{2}^{2} (\psi,\e, 0) {\sf D}_{2} \right)
\equiv Q^{-2} {\hat {\sf D}}_{-2} + Q^2 {\hat {\sf D}}_{2} \ \ ,
\ee 
where ${\sf D}_p$ define network vectors with ${D}^I_p$ as their 
components; ${\hat {\sf D}}_p$ are their normalized counterparts. Thus,
$Q_2 = {\hat {\sf D}}_{2}\cdot {\sf Q} = Q^{+2} + Q^{-2} {\hat {\sf D}}_{2}
\cdot {\hat {\sf D}}_{-2}$.
Hence, $\{{\hat {\sf D}}_{2}, \>{\hat {\sf D}}_{-2}\}$ define a two-dimensional
complex plane, ${\cal P}$ (the helicity space, a subspace of ${\cal C}^N$), 
on which a metric $g_{pq}$ can be defined. Then, $Q_p = g_{pq} Q^q$ with $p,q
=\pm 2$. The $N$-dimensional correlation vector ${\sf C}$, in general, lies 
outside ${\cal P}$. However, ${\sf Q}$ lies totally in ${\cal P}$. Thus, the 
statistic reduces to
\be
\ln\lambda |_{\hat{\sf b},\hat{\d}} = 
\left| {\sf Q} \cdot {\sf C}^* \right|^2 /2= \left| {\sf Q} \cdot 
{\sf C}^*_{\cal P} \right|^2 /2
\ \ , \ee
where ${\sf C}_{\cal P}$ is the projection of ${\sf C}$ onto the helicity 
space.

Maximization of $\ln\lambda |_{\hat{\sf b},\hat{\d}}$ over $\{\psi,\e\}$ is 
achieved by aligning  
${\sf Q}$ along ${\sf C}_{\cal P}$. This requires that $Q^{\pm 2} = 
{C^{\pm 2}_{\cal P} / {\parallel {{\sf C}_{\cal P}} \parallel}_{\cal P}}$, 
which implies:
${Q^{+2} / {Q^{-2}}} = 
{{C^{+2}_{\cal P}}/ C^{-2}_{\cal P}}$.
Since the RHS above can take any value in the complex plane, we need
to prove that ${Q^{+2} / {Q^{-2}}}$ obeys the same property. To this end, 
note that,
\be
{Q^{+2} \over {Q^{-2}}} = 
{{T_{2}^{-2} (\psi,\e, 0)} \over {T_{+2}^{2} (\psi,\e, 0)}} = 
\left ({{1 - \cos \e} \over {1 + \cos \e}}\right)^2 \exp({4i \psi}) \ \ ,
\ee
which can indeed attain any value on the Argand plane. The values of $\psi$
and $\e$ that 
maximize the statistic are, 
${\hat \psi} = \arg (x)/4$ and
${\hat \e} = \cos^{-1} [(1-\sqrt{\|x\|}\>)/(1+\sqrt{\|x\|}\>)]$ 
where $x \equiv C^{+2}_{\cal P} / C^{{-2}}_{\cal P}$. 
Thus, the statistic maximized over these four parameters is, 
\be
\label{st}
\ln\lambda |_{{\hat {\sf b}},{\hat{\d}},{\hat \psi},{\hat \e}} = 
{\parallel {\sf C}_{\cal P} \parallel}^{2} /2\,.
\ee
Let ${\hat V}^\pm$ denote a pair of orthonormal, complex basis vectors 
on ${\cal P}$. Then,
\be
\label{st2}
{\parallel {\sf C}_{\cal P} \parallel}^{2} = {\parallel {C}^{+} 
\parallel}^{2} +{\parallel {C}^{-} \parallel}^{2} 
 = (c_0^+)^2 + (c_{\pi/2}^+)^2 + (c_0^-)^2 + (c_{\pi/2}^-)^2 \ \ ,
\ee
where $C^\pm = {\sf C}_{\cal P} \cdot {\hat V}^\pm =c_0^\pm + ic_{\pi/2}^\pm$. 
It can be verified that the 
network statistic is, therefore, a sum of the squares of four Gaussian random 
variables with constant variance. This simplifies the computation of 
thresholds and detection probabilities. 

A network filter is constructed as follows: For a given $\xi$ compute the 
Newtonian chirp for $t_a = 0$. Then, for a given direction $\{\phi,\theta\}$,
use the appropriate time-delays with respect to fide to time-displace the 
chirp at each detector. This collection of time-displaced chirps 
constitute the network filter. Also, ${\hat{t}}_a$ is obtained
by shifting the network filter `rigidly' on the time axis, which can be done 
efficiently using FFT. The bank of filters on $\{\phi,\theta,\xi\}$ can be 
obtained by correlating two neighboring normalized filters
in the usual way. This work is now in progress. 

\section{Conclusion}

We have given here a formulation to optimally detect the Newtonian chirp with 
a network of detectors in which the noise is additive, Gaussian, and 
uncorrelated between detectors. We have shown how this can be done efficiently
by analytically maximizing the LR over four parameters and using FFT to  
maximize over the time-of-arrival. In a future work we hope to address key
issues such as required computational power for such a search and also 
estimate errors in parameter values. 

\acknowledgments

SB acknowledges support from PPARC grant PPA/G/S/1997/00276. 
AP acknowledges support from a CSIR grant and from AEI, Potsdam, for a 3-month 
visit.


\begin{references}
\bibitem{6} 
In Ref. [2], $W^{(I)}$ is the symbol used 
for denoting a quantity proportional to $F^I$.
\end{references}
\end{document}